\documentstyle[12pt,a4]{article}
\begin{document}

\def\msbar{{\rm \overline{MS\kern-0.14em}\kern0.14em}}

\begin{flushright}
 Liverpool Preprint: LTH 279\\
January 24, 1992 Revised
\end{flushright}
\vspace{5mm}
\begin{center}

{\LARGE\bf The Running Coupling from Lattice Gauge Theory}\\[1cm]

{\bf C.\ Michael}\\
 DAMTP, University of Liverpool, Liverpool L69 3BX, U.K.

\end{center}

\begin{abstract}

 From an accurate determination of the inter-quark potential, one can
study the running coupling constant for a range of $R$-values
and hence estimate the scale $\Lambda_{\msbar} $.
Detailed results are  presented  for $SU(2)$  pure  gauge  theory  to
illustrate the method.
\end{abstract}

\section{Introduction}
\par
In the continuum the potential between static quarks is known
perturbatively to two loops in terms of  the  scale $\Lambda_{\msbar} $.
 For  $SU(2)$ colour, the continuum force is given by \cite{bill}
$$
{dV \over dR } =  {3 \over 4} {\alpha(R) \over R^2}
$$
\noindent with the effective coupling $\alpha (R)$ defined as
$$
\alpha (R) = { 1 \over 4 \pi [ b_0  \log (R\Lambda _R )^{-2} +
(b_1 / b_0 ) \log \log (R\Lambda_R )^{-2} ] }
$$
\noindent where $b_0=11/24 \pi ^2$ and $b_1=102 \ b_0^2/121$ are
 the usual coefficients in the
perturbative expression for the $\beta$-function
and, neglecting quark loops in the vacuum,  $\Lambda_R= 2.055
\Lambda_{\msbar}$.
Note that the usual lattice regularisation scale $\Lambda_L = 0.05045
 \Lambda_{\msbar}$.
\par
 At large separation $R$, the potential behaves as $KR$ where $K$  is  the
string tension.  Thus in principle knowledge of  the  potential $V(R)$
 serves  to
determine the dimensionless ratio $\sqrt K/\Lambda $ which relates the
perturbative scale $\Lambda$ to a non-perturbative observable such as the
string tension $K$.  This is the  basis  of
the method we shall employ here.
\par
It is worth recalling the lattice method  that  has  been  used
previously: namely determining the dimensionless
string tension $Ka^{2}$ directly  from
the large-$R$ potential and then using the  perturbative  relationship
between $a$ and $\beta $ to find $\sqrt K/\Lambda $.  This method requires
  that $a(\beta )$
is given by the two  loop  perturbative  beta-function:  a  condition
known as asymptotic scaling  in  the  lattice  gauge  theory  realm.
Ample evidence exists  that  this  condition  is  not  satisfied  at
present $\beta $-values (up to $\beta =2.85$ for $SU(2)$  pure  gauge
 theory  for
example the beta-function is only $82\%$ of the perturbative value
 \cite{ukqcd}).
\par
However, though asymptotic scaling is  not  yet  manifest,  the
weaker  scaling   requirement   is   well   satisfied.    Thus   the
dimensionless  ratios  of  physical  quantities  are  found  to   be
independent of $\beta $.  For example the  potential $V(R)$  scales
\cite{ukqcd}  over  a
range of lattice spacing of a factor of 4  (from $\beta =2.4$  to $ 2.85$).
That scaling but not asymptotic scaling is valid, implies  that  the
bare coupling constant derived from $\beta $ is inappropriate and  that  an
effective coupling constant derived from some physical quantity is a
better choice.  This has been emphasized by  Lepage  and  Mackenzie \cite{lm}.
It is also the basis of the method proposed by L\"uscher et al~\cite{lu}
 to extract the running coupling constant.
Here we  use  lattice  simulation  to  determine  the  interquark
potential between static quarks and  so  obtain  the
running coupling constant at small distance $R$.
 One subtlety is that we require small $R$ and hence large energy $1/R$
to make most precise contact with the  perturbative  expression  but
the lattice method implies the presence of lattice artefacts when $R
\approx  a$.  Our main concern will be  to  show  how  to  cope  with  these
lattice artefacts  and  the  method  will  be  to  study  potentials
off-axis on the lattice as well as on-axis.
\par
\section{Lattice potentials}
\par
To explore the interquark potential precisely, we use as large
 a lattice as feasible and as large  a  value  of $\beta $  consistent  with
remaining in the large-volume vacuum.  Previous work \cite{pm} has shown  that
$\beta =2.7$ and a $32^{4}$  lattice  is  suitable.   We  use  rather  similar
methods to  those  used  previously to determine the string tension
\cite{pm}  but  here  we  concentrate  our
attention on small $R$.   At  small
$R$, the  statistical  errors  will  be  rather  small  and  the  main
uncertainty will be  the  systematic  effects  coming  from  lattice
artefacts.    Thus   we   measure   the   potentials   at  on-axis
separations with $R/a$ =1, 2, 3, 4, 6, 8, 10, 12, 14, 16 and off-axis
separations with $R/a$ vectors (1,1,0), (2,1,0), (2,2,0), (3,1,0), (3,2,0)
 and (3,3,0).
\par
 The method used to extract the  potentials  is  to  use
spatially smeared links  (APE  smearing \cite{ape}   with $SU(2)$
  projection  of
$2.5 \ \times$  straight link plus four  spatial U-bends)  to  build  up  paths
between the static sources.   Recursive  smearing  with  30  and  60
levels is used, so providing two paths which gives a $2\times 2$ correlation
matrix that can be used in the standard  variational  approach. For the
on-axis separation we use straight paths, while for the off-axis
separations we sum over two L-shaped paths.  We
measure potentials for the $R$-separations above and $T$-separations 0 to 5.
The lattice is well equilibrated from previous work and  we  measure
40 blocks of 6 configurations with each block containing 150  update
sweeps (3 over-relaxation to 1 heat bath).
\par
The potentials are given by the extrapolation in $T$ of the ratio
of generalised Wilson loops.
$$
   V(R) = \lim_{T \to \infty} V(R,T),
$$
\noindent where
$$
   V(R,T)= -\log W(R,T)/W(R,T-a)
$$
\noindent  This is a  monotonic  decrease with $T$  and  the  rate  of
decrease can be estimated from the energy  gap  between  the  ground
state (which is what we wish to determine)  and  the  first  excited
state.  We obtain estimates of this energy gap from our  variational
method in the 2:1 $T$-ratio basis and these estimates agree
 with previous work \cite{pm}.  We then   use
those  estimates  to  complete   the $T$-extrapolation.    We   find
consistency between such extrapolations based on $T$-values 2-4 and on
$T$-values 3-5 which confirms the  stability  of  the  method.   Error
analyses use bootstrap from our 40  samples  which  we  find  to  be
consistent with being statistically independent. These results for
the potentials agree within 1 $\sigma$ with those of ref.\cite{pm} at
common $R$-values.  The force derived from our potential measurements
is shown in table 1.
\par
\begin{table}
\begin{center}
\begin{tabular}{|r|l|c|l|}\hline
 $R/a$ \  & $\ \Delta V/\Delta R $  & $\Delta V_c / \Delta R$ &
 $ \qquad \alpha(R)$  \\\hline
 1.2071 & 0.1745(2)    & 0.1339    & 0.2525(4)(76)    \\
 1.7071 & 0.0640(2)    & 0.0798    & 0.3009(8)(60)    \\
 2.1180 & 0.0829(6)    & 0.0570    & 0.3401(35)(154)   \\
 2.5322 & 0.0463(3)    & 0.0448    & 0.3774(28)(13)   \\
 2.9142 & 0.0203(14)   & 0.0345    & 0.3909(153)(161)  \\
 3.0811 & 0.0437(10)   & 0.0347    & 0.4384(132)(114)  \\
 3.3839 & 0.0321(5)    & 0.0304    & 0.4616(80)(27)   \\
 3.8028 & 0.0234(7)    & 0.0255    & 0.4913(144)(42)  \\
 4.1213 & 0.0294(11)   & 0.0254    & 0.5740(242)(90)  \\
 5.0000 & 0.0207(2)    & 0.0204    & 0.6526(57)(10)   \\
 7.0000 & 0.0154(2)    & 0.0153    & 0.9797(121)(4)  \\
 9.0000 & 0.0132(2)    & 0.0132    & 1.4128(215)(2)  \\
11.0000 & 0.0128(4)    & 0.0128    & 2.0408(571)(1)  \\
13.0000 & 0.0117(3)    & 0.0117    & 2.6124(680)(1)  \\
15.0000 & 0.0112(3)    & 0.0112    & 3.3546(973)(1)  \\\hline
\end{tabular}
\caption{ The force $\Delta V/\Delta R$ and lattice artefact corrected force
$ \Delta V_c / \Delta R $ at
average separation $R$. The running coupling $\alpha (R)$ derived
from the corrected force is shown as well. The second error shown on
$\alpha$ is $10\%$ of the lattice artefact correction.}
\end{center}
\end{table}
The potential shows a lack of rotational invariance at small $R$.
To lowest order this can be attributed to the difference $\delta G(R)$ between
the lattice one gluon exchange expression and the continuum expression.

$$
\delta G(R) =  {4 \pi \over a} \int_{-\pi}^{\pi}
 { d^3 k \over(2\pi)^3}
{ e^{ik.R/a} \over 4\sum_{i=1}^3 \sin^2 (k_i / 2) }
-{1 \over R}
$$
On a lattice, the next order of perturbation has been calculated \cite{hk}
and the dominant effect is a change from the bare coupling
to an effective coupling \cite{lm}. In that case,  using the difference
above but with an adjustable strength will correct for the small $R/a$
lack of rotational invariance. A test of this will be that a smooth
interpolation of $V(R)$ versus $R$ is obtained with this one free
parameter to the 6 off-axis potential values.

  We  evaluate $\delta G(R)$
numerically using the limit of a very large lattice since we are not
here concerned with long-range effects.  Then we find the  following
empirical expression provides a good fit for $R~> a$,
$$
aV(R)=C- {A \over R}+{B \over R^2}+KR - A f \delta G(R),
$$

\noindent with $\chi ^{2}$ per degree of
 freedom 0.8 using the full correlation matrix. The fit parameters are
shown in table 2. For our present purposes
the detailed form of this fit at small $R$ is not relevant
 - what is needed is a confirmation that a good fit
can be obtained. This then supports  our prescription to
 correct the lattice artefacts responsible for the lack of rotational
invariance. What is more difficult is to assign errors to this
correction procedure. Since the correction coefficient $f$ is determined
to a few per cent, the statistical errors are small. The fact that one
parameter corrects 6 off-axis points simultaneously is very
encouraging. The only way to be certain that lattice artefacts are
eliminated is by the comparison
of different $\beta$ values (with thus different $R/a$ values at the
same physical $R$ value) and we shall see later that this test is satisfied.
 This leads us to use as an illustration a $10\% $ systematic error to the
 correction itself
(3 times the statistical error) with the proviso that
 for the lowest $R$ value ( $R=a$ ) the smooth interpolation is less of a
constraint so that we disregard that datum in the analysis.
We then assume that an improved estimate for the continuum potential $V_c$ will
be obtained by correcting the measured lattice values $V$ by $\delta G$ with
the
fitted coefficient.  These values are  shown  in table  1.

\begin{table}
\begin{center}
\begin{tabular}{|c|c|c|c|}\hline
$A$ & $f$ & $Ka^2$ & $B/a$ \\\hline
 0.261(5) & 0.68(3) & 0.0103(2) & 0.054(5) \\\hline
\end{tabular}
\caption{Fit to force for $R>a$.}
\end{center}
\end{table}
\par
It is now  straightforward  to  extract  the  running  coupling
constant by using
$$
\alpha( { R_1 + R_2 \over 2 }) = { 4\over 3} R_1 R_2 { V_c(R_1)-V_c(R_2) \over
 R_1-R_2 }
$$
\noindent where the error in using a finite difference is here negligible.
This is shown in table 1 and is plotted  in the figure versus
$R/ \sqrt K$ where $K$ is taken from the fit - see Table 2.
The interpretation of $\alpha$ as defined above as an
effective running coupling constant is only justified at small $R$ where the
perturbative expression dominates.
  Also shown  are  the
two-loop perturbative results for $\alpha(R)$ for
different values of $\Lambda_L $.
\begin{figure}
\vskip 4.5in
\caption{
The effective running coupling constant $\alpha(R)$ obtained from
the force betwen static quarks at separation $R$.  The scale is set
 by the string tension $K$.  The dotted error bars represent an estimate
of the systematic error due to lattice artefact correction as described
in the text.  The curves are the two-loop perturbative expression.
}
\end{figure}
\par
The figure clearly shows a {\it running} coupling constant.  Moreover
the result is consistent with the expected perturbative dependence  on $R$  at
small $R$.  There are systematic errors, however. At larger $R$, the
perturbative two-loop expression will not be an accurate estimate
of the measured potentials, while at smaller $R$, the lattice artefact
corrections are relatively big.  Setting the scale using $\sqrt K=0.44$ GeV
implies $1/a(2.7)=4.34 $ GeV, so $R < 4a(2.7)$ corresponds to
values of $1/R > 1$ GeV.  This $R$-region is expected to be adequately
 described
by perturbation theory.  Another indication that perturbation
theory is accurate at such $R$-values is that $\Delta V_c / \Delta R$
at small $R$ is found to be very much greater than the non-perturbative
 value $K$ at large $R$.
\par
The best way to gain confidence that
 these systematic errors are under control, is to repeat the method
at another $\beta$-value.
  The UKQCD data \cite{ukqcd} at $\beta =2.85$ for the potential at
 $ R/a(2.85)$ =2, 4 and 6  are processed in the same  way to yield
$\alpha(R)$ at $R/a(2.85)$=3 and 5.  Since only the on-axis values
 were measured, we fix $fA$ at
the value found in the above fit at $\beta=$2.7
for the lattice artefact correction.  Since we have no cross-check from
off-axis potentials,  we assign
a larger systematic error $(30\%)$  to  this  lattice  artefact
correction.  The results for the effective running coupling
are shown in the figure and are seen to confirm nicely the result
 from $\beta =2.7.$
\par
The easiest way to describe the value of the running coupling constant
$\alpha$
is in terms of a scale or $\Lambda$ value with the understanding that
we are only determining $\alpha$ for a range of energy scales
 $1/R$ - namely 1 to 3 GeV.
The final estimate of $\Lambda$ is made from the figure, weighting
smaller $R$ more heavily since the perturbative expression is
more accurate as $\alpha(R)$ becomes smaller. We exclude the lowest
$R$ point since the lattice artefact correction for $R=a$ is
untested.  Remembering that the systematic errors due
to lattice artefact correction are estimates only and since these systematic
errors are dominant, we do not attempt a fit but we can  conclude
that our results are consistent with values of
 $\Lambda$ lying in the range shown by the two curves plotted.
{}From the data at $\beta $ = 2.7, these curves have
$a(2.7)\Lambda_R$=0.0619 and 0.0688.  Using the value of the string
tension from the fit, we get $\sqrt K/\Lambda_L$= 31.9(1.7).  Moreover, this
value is consistent with the evaluation at both $\beta=$ 2.7 and 2.85.

\par
\section{Conclusions}
\par
Using  the bare
coupling $g$ derived from $\beta =4/g^{2}$ and the  two-loop  perturbative
 relationship  $a(g)$ in terms of the scale $\Lambda_L$
gave \cite{pm,ukqcd} the following
slowly decreasing values of $\sqrt K/\Lambda_L =$ 53.3(3), 49.1(4)and 44.1(6)
 at $\beta =$2.5, 2.7 and 2.85 respectively.  Clearly, the
$\beta \rightarrow \infty $ limit lies below these values.
 Our present method which does not rely on the bare coupling,
 gives  the scaling result which should be independent of $\beta$.
Our estimate is  $ \sqrt K /\Lambda_L = 31.9(1.7)$.    This
is sufficiently far below the values extracted from the bare coupling
to imply  that asymptotic scaling to two-loop perturbation theory is
 not ``just around the corner'' but will only  be
satisfied accurately at larger $\beta $-values than those currently
accessible to lattice simulation.
\par
Our method corresponds to an
estimate of  the continuum ratio $\sqrt K / \Lambda_{\msbar}$ =
1.61(9) for pure SU(2) theory. Setting the scale using $\sqrt K$ = 0.44
GeV, then gives $\Lambda_{\msbar}$=273(15) MeV.   These results are
obtained for rather modest energies ( $1/R \approx 1 - 3 $ GeV ) and
it is important to extend the lattice methods to higher energies too.
 From lattice results for ratios of other non-perturbative
quantities (glueball masses, critical temperature, etc) to the
string tension, one can then determine their value in terms of
$\Lambda_{\msbar}$ as well.
\par
As well as the case of $SU(2)$, we  can  apply  the  same  method
directly to $SU(3)$.  Using published data \cite{su3} at
 $\beta =6.2$ gives an  estimate
of $\Lambda_{\msbar} = 250-300$ MeV. In order to improve on this
 determination, it
will be necessary to study the small$-R$ on-  and  off-axis  potential
accurately at larger $\beta $ for $SU(3)$.  This method will  then  determine
the  running coupling accurately in  terms  of  any  other  physical
quantity measurable on the lattice (such as the string  tension $K)$.
This  lattice  method  gives  an  accuracy  which  is competitive with that  of
experimental determinations of the running coupling  for modest energy
scales. The difference, however, is that these
lattice methods are feasible for pure gauge simulation but we have yet to
achieve similar results for full QCD.
\par
\bigskip
\noindent I wish to acknowledge the suggestion of  Rainer Sommer
who emphasized to me the feasibility of extracting the running
coupling  from the small-$R$ potentials.

\end{document}